"A fresh look at some questions surrounding black holes"

M. Anyon and J. Dunning-Davies,

Department of Physics,

University of Hull,

Hull  HU6 7RX,

England.

email: j.dunning-davies@hull.ac.uk

**Abstract**

The modern notion of a black-hole singularity is considered with reference to the Schwarzchild solution to Einstein's field equations of general relativity. A brief derivation of both the original and the modern line elements is given. The argument is put forward that the singularity occurring within the Schwarzchild line element, that has been associated with the radius of the black-hole event horizon, is in fact merely a mathematical occurrence, and does not exist physically. The real aim here, however, is to attempt to open up the whole problem, draw some conclusions but finally to urge everyone to consider the points raised with no preconceived opinions and then come to their own final conclusion.

One of the most fascinating and mysterious concepts within physics is the ever intriguing notion of a black-hole. Beloved by science fiction writers, readily accepted, and misunderstood, by the public and fervently investigated by many physicists, including the élite of theoretical physics and cosmology, these elusive objects certainly capture the imagination. But what exactly is a 'black-hole'?

It seems that the term was first coined by John Wheeler, and modern-day black holes are heralded as consequences of Einstein's general relativity. They are commonly assumed to be points in space where all the known laws of physics break down and all matter is compressed down to an infinitely dense point. This view is mainly due to the singularity occurring within Karl Schwarzchild's solution to the field equations of general relativity of 1916. This solution explains the space-time about a body of given mass and has been used extensively.

A singularity, as defined by the Oxford Dictionary of Physics [1], is 'a point in space-time where general relativity theory predicts that certain quantities including the curvature of space-time become infinite'. Quite a succinct, and possibly too simplistic, explanation, but one that would probably define most people's understanding of the object in question. However, when one considers what this actually means it becomes apparent that this definition is merely attaching a physical explanation to a mathematical expression at the point where the model breaks down and becomes infinite. The four dimensional space-time of general relativity is simply a mathematical space used to try to explain properties of the universe. It is not what the universe actually looks like. This has to be confirmed observationally and experimentally in order to maintain the synchronicity between the real world and the results of the model. Infinities are usually thought to be the limiting point of a mathematical model and either indicate that the model is incorrect or incomplete, or merely a consequence of the mathematical model itself and cannot be attributed to any physical process or object. Attaching the physical interpretation that the mathematical Schwarzchild singularity is a black-hole in physical space may well have led the scientific community down a long road that could lead nowhere. Loosely defined, the idea of a 'black-hole' can arise in two quite distinct ways.

Yet they are linked quite nicely by what could only be considered an amazing coincidence.

One such candidate for a black-hole is a body described by John Michell in 1784 [2]. Michell derived an expression giving the ratio of the mass to radius of a body with an escape velocity equal to or greater than the speed of light, *c*, given by

$$M/r \geq c^2/2G = 6.7 \times 10^{26} \text{ kg/m}$$

Similarly Laplace in 1796 published a proof of the result that a sufficiently massive and condensed body could appear invisible. [3] However, such a body might better be described as a dark body. It would, after all, merely be a very dense body which could be approached and, in fact, viewed from a suitable distance, unlike the popular notion of a black hole.

On the other hand, a relativistic 'black-hole' occurs as a consequence of the singularity appearing within the Schwarzschild solution to Einstein's field equations of general relativity for a spherically symmetric mass point [4]. In this solution it was noticed, however, that a singularity appeared when the mass and radius of the event horizon were related in the same way as in Michell's result above, given approximately 130 years previously! This Schwarzschild singularity became associated with the end point of the life of a massive star in the 1960s following work by Oppenheimer and associates, and the modern notion of a black-hole was born, with general relativity apparently predicting that at the centre of a black-hole is a singularity where density becomes infinite and the known laws of physics break down. [5] But why has this viewpoint persisted?

By the 1960's it was generally accepted that, when a star has used all of its thermonuclear energy, it contracts under its own gravity and forms a degenerate star. At extremely high densities, the collapse is halted by the electron or neutron degeneracy pressure forces. Star cores below approximately one solar mass, $1.4 M_\odot$, contract to what are known as white dwarfs, comprised of a degenerate electron gas. The electrons provide a very strong resistive pressure due to the fact that they are fermions. This means that they obey the Pauli Exclusion Principle and, therefore, under great compression, cannot exist in the same space and energy state at the same time as each other. This pressure force cannot hold back the collapse of dying star cores of from about 1.4M$_\odot$ up to about 3M$_\odot$ though. These cease their collapse with the formation of a neutron star, comprised of a degenerate neutron gas . However, stars of mass greater than approximately 3M$_\odot$, were thought to contract indefinitely, since the force of gravity would overwhelm all other forces. The problem of what would happen to a collapsing star of mass greater than about 3M$_\odot$ was considered by R. Oppenheimer and H. Snyder in 1939 [6]. They stated that general relativity was needed to describe the dynamics of the star's collapse [3] and so based their work on Schwarzchild's solution from 1916. Since Schwarzchild's solution is a description of the space-time metric about a gravitating body of a particular mass it was ideal for this problem.

In most standard textbooks on the General Theory of Relativity [7] Schwarzchild's solution of the Einstein field equations is stated as being either;

$$ds^2 = \left\{1 - \frac{2Gm}{rc^2}\right\}c^2 dt^2 - \left\{1 - \frac{2Gm}{rc^2}\right\}^{-1} dr^2 - r^2\left(d\theta^2 + \sin^2\theta d\phi^2\right)$$

or

$$ds^2 = \left\{1 - \frac{2m}{r}\right\}dt^2 - \left\{1 - \frac{2m}{r}\right\}^{-1} dr^2 - r^2\left(d\theta^2 + \sin^2\theta d\phi^2\right)$$

[8], with c = G = unity for the latter expression, and r, $\theta$ and $\phi$ being the spherical polar co-ordinates for both.

The solution shows that using general relativity, the space-time metric about a spherical mass exhibits singular behaviour at a radius, known as the Schwarzchild radius;

$$R_s = 2GM/c^2$$

where G = Newton's gravitational constant, M = mass of the star and c = speed of light. This relation also leads to exactly the same ratio of mass to radius as derived by Michell in 1784 of

$$M/R_s = c^2/2G = 6.7 \times 10^{26} \text{ kg/m}$$

Using this result, Oppenheimer and Snyder demonstrated that if the mass of a collapsing star is encompassed within this radius it will collapse in an infinite time within this radius, as observed by an external observer. This radius would then define a trapped region of space from which nothing can escape, and is completely inaccessible from the outside [3]. So, theoretically, such a body could exist in both Newtonian and relativistic mechanics, but in relativity theory the 'body' with an escape speed greater than *c* becomes a singularity.

Tassoul and Tassoul's coverage of the subject of black-holes in their book, 'A Concise History of Solar and Stellar Physics' [3], does accept that such a physical body could exist, but mentions little of the nature of the singularity: 'Although these bounds (on mass and radius) might suggest the existence of a body that has the properties of a black-hole, it is important to recall that the physics of ultra dense matter beyond the neutron-star stage is still poorly understood' [3]. As shown above, the existence of a body with an escape speed greater than or equal to the speed of light is quite possible in Newtonian mechanics, as shown by Michell, and Laplace. So, what is the nature of the Schwarzchild singularity within general relativity?

In order to derive the general relativistic line element, it is necessary to obtain the field equations. It is then possible to derive a metric expression for the line element *ds*. This metric must be the same for all points located at the same radial distance from the central mass source, assuming it to be spherically symmetric. For this spherical polar co-ordinates are required. A derivation of the Schwarzchild metric leads to

$$ds^2 = (1-2m/r)\,dt^2 - (1-2m/r)^{-1}\,dr^2 - r^2\,(d\theta^2 + \sin^2\theta\,d\varphi^2) \,, \tag{a}$$

which is the usual form when taking $c = G = 1$, and is known as the Schwarzchild line element.

In this line element, a mathematical singularity occurs when r = 0 as is to be expected due to the co-ordinate system employed. However, due to the form of the coefficient $dr^2$ it is seen that a second singularity appears when $r = 2m$. This is the singularity which has been explained by associating it with the existence of black-holes. It is quite easy to see that the expression exhibits singularities at both these points, but the explanation of the second singularity is the focus of this discussion. The Schwarzchild solution is considered in detail in the 'General Theory of Relativity' by P.A.M. Dirac [9] and the singularity at $r = 2m$ is analysed. It is found that a particle close to the critical radius will take an infinite time to cross the threshold as seen by an observer at infinity. Considering a light emitting particle near the point $r = 2m$, it is found that the red-shift of the light also tends to infinity as the critical radius is reached, meaning that the body in question is inaccessible to the outside universe. This agrees with the conclusion of Oppenheimer and Snyder as stated above [6]. With regard to the matter falling towards this boundary it is taken that the prior interpretation of the solution is that $r = 2m$ defines a minimum radius for a body of mass $m$, but after closer consideration and investigation it is found that this is not so. Consider an observer travelling with the matter, this observer will have aged only a finite amount when $r = 2m$ is reached. It is, therefore, required to examine the line element for values of $r \leq 2m$. Dirac is able to continue the solution to values of $r \leq 2$m by using a non-static co-ordinate system in order that the metric tensor varies with time. This is achieved by retaining $\theta$ and $\phi$ as co-ordinates but defining $t$ and $r$ by;

$$\tau = t + f(r) \quad \text{and} \quad \rho = t + g(r),$$

where the functions $f$ and $g$ are freely at the disposal of the investigator. [8, 9]

A Schwarzchild line element is found using the new co-ordinate system and the critical radius $r = 2m$ is found to correspond to a value of $\rho - \tau = 4m/3$, where there is no singularity. Dirac notes that the new line element holds for regions $r > 2m$ because it can be transformed to the original solution by a change of co-ordinates, and it also holds for values of $r < 2m$ since there is now no singularity occurring at $r = 2m$. This demonstrates quite clearly that the singularity in question is purely a mathematical one, not a physical one, and only occurs at the point of connection between the two co-ordinate systems. The region $r < 2m$, however, cannot communicate with regions of $r > 2m$ as any signal takes an infinite time to cross this boundary as measured by external clocks. Dirac continues by asking the question of whether or not such an object can exist in reality. The only thing that can be said with any certainty, according to Dirac, is that the Einstein equations allow it. 'A massive stellar object may collapse to a very small radius and the gravitational forces then become so strong that no known physical forces can hold them in check and prevent further collapse' [9]. The discussion then goes on to say that if this were to happen it would then seem, logically, that the matter would *have to* collapse into a black-hole. This seems like an ambiguous comment following the discussion given, and it appears as if Dirac might be playing safe and not committing himself to either side of the argument. This viewpoint is one which Einstein did not ascribe to, however. In fact he devoted an entire paper, in 1939 [10], to discussing the nature of the Schwarzchild singularity and whether or not it existed in reality. Einstein's conclusion was that it did not, and the Schwarzchild singularity was a mathematical notion alone; any evidence for such a body would have to come from direct experimental and observational data. Amazingly, the 'father of black holes', denied their existence from the outset. The classical viewpoint, however, of Michell and Laplace, of a body so dense that the escape velocity exceeds the speed of light, *c*, appears to be a lot more plausible. If special relativity holds at such a point, and this ratio is exceeded, then the apparent time dilations would effect the observations in the way predicted by Oppenheimer, Snyder and Dirac, and the body would, to all intents and purposes, be an 'inaccessible dark star'. However, the idea that such a body would create a singularity seems unacceptable, particularly since the above arguments suggest that the idea of a point singularity is entirely mathematical in foundation.

The discussion so far, however, has only covered the line elements that are commonly given in modern text books and scientific papers. The original solution was published almost 100 years ago, so to ignore it would be careless. Modifications and alterations for simplicity are bound to occur over that period of time, and misconceptions can easily arise. So to fully understand the solution in its original context we must consider the original publication. This paper, published in German in 1916, has recently been carefully translated by S. Antoci and A. Loinger making it more accessible to the general scientific community. [4] The paper is a very thorough and exact solution to the problem of describing the gravitational field of a symmetrically spherical mass point according to Einstein's general relativity. As will be explained, Schwarzchild considered the notion of what he calls a 'discontinuity' in the line element, and solves the problem after a discussion about the physical interpretation of the line element. This rigorous solution therefore, as described in the forward by its translators, 'leaves no room for the science fiction of black holes.' [4]

The derivation is very similar to the one given above, but with different notation and one vitally important difference. Schwarzchild lists a set of requirements that must be fulfilled in order for the solution to hold. The form of the expression derived must be such that the field equations remain unaltered when calculating it in polar co-ordinates, consistent with the principle of covariance. This is discussed by Schwarzchild who states that in order for this to be the case; the field equations (must) have the fundamental property that they preserve their form under the substitution of other arbitrary values in lieu of $x_1$, $x_2$, $x_3$, $x_4$, as long as the determinant of the substitution is equal to 1. To derive the solution, and retain the determinant equal to 1, the three orthogonal spatial components are defined as:

$$x_1 = r^3/3 \qquad x_2 = -\cos\theta \qquad x_3 = \phi$$

The new variables are then polar co-ordinates with the determinant equal to 1. The solution is then derived by substituting the variables into the general form of the line element, taking note of the conditions which must be fulfilled. These include:

The equation of the determinant must equal unity

The field equations must be satisfied

Continuity of the function, except for $x_1 = 0$

[4]. The line element is then derived in the form:

$$ds^2 = (1-\alpha/R)\, dt^2 - (1-\alpha/R)^{-1} dR^2 - R^2 (d\theta^2 + \sin^2\theta d\varphi^2)$$

where the quantity $R = (3x_1 + \rho) = (r^3 + \alpha^3)^{1/3}$ and $\rho$ is an arbitrary constant. It should be noted that the constant $\alpha$ depends only on the mass of the body at the origin, and can be compared to the constant $2m$ in equation (a).

Schwarzchild goes on to discuss this line element, and states that without the condition of continuity the solution would be of a different form entirely. The exact form of the solution shows that the discontinuity occurs at a value of $r = (\alpha^3 - \rho)^{1/3}$, and therefore one need simply set the arbitrary constant $\rho = \alpha^3$ for the discontinuity to go to the origin. Analysing the above line element shows that it exhibits singular behaviour at $R = \alpha$, which appears very similar to the situation of $r = 2m$, obtained using equation (a). However, it should be noted that when $R = \alpha$, the radial distance component $r = 0$, and so the singularity occurs at the origin.

This demonstrates clearly that the original Schwarzchild solution to Einstein's field equations for a spherically symmetric mass source does not contain a second singularity that can be attributed to the minimum radial distance of a gravitating body - leading to the formation of a singularity. Although it is true that, if $R = \alpha$, the solution will show singular behaviour, this value of $R$ is not the radial spatial component associated with polar co-ordinates. This fact, however, has been repeatedly ignored, and claims of black-hole discoveries are pronounced constantly.

Observational evidence for black-holes provided in the numerous papers and letters claiming their existence is sketchy at best, even though there are constantly new 'sightings' of black-holes claimed by scientists. Indeed the new fad at the moment is the quest for the 'super-massive black-holes' at the centre of galaxies. These claims of 'sightings' though are via entirely indirect methods - since a black-hole is an object with an escape speed greater than the speed of light and therefore, cannot be observed directly. These observations are based on high intensity radiation emissions from the central parts of galaxies. These are obviously very dense regions of stellar matter, and appear extremely bright throughout the electromagnetic spectrum. It is reasoned that the matter falling in towards the event horizon of a black-hole emits large amounts of radiation due to the fact that this matter is being accelerated by gravity towards the singularity. This acceleration, as well as the collisions of particles within the accretion disk, cause electromagnetic radiation of extremely high energy to be emitted, and intense X-ray emissions are often cited as evidence for black-holes. However, these emissions could be explained in a number of different ways, and the criteria necessary to definitively confirm the existence of a black-hole - the mass to radius ratio given by Michell and the existence of an event horizon - are yet to be met! At no point in the field of black-hole physics has a candidate singularity met these criteria, yet the announcement of the discovery of bigger and more powerful black-holes comes ever more frequently.

The idea of a singularity occurring within the centre of a collapsed star was not attributed to the Schwarzschild line element until approximately the 1960s, when work by Oppenheimer, Wheeler and associates on the end-point to the lives of stars was being carried out. A return to the original term, 'Dark Stars', coined by Michell in the 18$^{th}$ Century would probably dispel some of the confusion regarding the nature of these objects, and would probably be more appropriate. The idea of an extremely large amount of matter being compressed to an infinitesimally small point with infinite density, a singularity, seems counter-intuitive and un-physical to some. The notion of a 'Dark Star', however, does not rely explicitly on the framework or foundations of general relativity and was proven to be possible, at least theoretically, by Michell as far back as 1784! It may be prudent to consider the following passage from Robert Wald in his book 'General Relativity';

'The above predictions of singularities in cosmology and gravitational collapse are based solely on the analysis of solutions with a very high degree of symmetry. It certainly is possible that they could give a completely misleading picture of singularity formation. For example, in the Newtonian theory of gravity, if a spherical, non-rotating shell of dust is released from rest, a singularity will be produced at r = 0 when all the matter simultaneously reaches the origin. However, if one perturbs the shell away from spherical symmetry or gives it some rotation, then no such singularity will occur.'

This seems to illustrate a point that needs emphasizing to many. The equations derived regarding the gravitational collapse of matter in space, such as the end life of a star or the large scale accretion of matter, do not actually describe the physical bodies with which they are associated. The Schwarzchild line element, for example, deals with the ideal situation of a spherically symmetric point mass source, whereas the Sun is in fact a non-symmetrical spherical rotating body with large finite radius. Although the practice of considering situations in their 'ideal' form for analysis is commonplace in physics, and the mass of the Sun can be considered to be a point source without affecting the results too adversely, these results should be considered 'idealized' approximations at best. When the above notion is considered with reference to black hole formation it seems that, in reality, the situation may be quite different from that given by the equations. If a star is undergoing gravitational collapse at the end of its life, it would be highly unlikely that the compression and collapse would occur completely homogenously and isotropically, so, certainly in Newtonian mechanics and quite possibly in general relativity, a singularity at $r = 0$ will not occur. The mathematics may allow it but the actual physical reality may well not correspond exactly to those mathematical results. Considering this point, and the fact that the second singularity occurring at the 'Schwarzchild radius' is entirely mathematical, it should become clear that the general perception and opinion of many possibly does not correspond to the reality.

Black-hole theory is based upon general relativity, and is derived from the Schwarzchild solution to Einstein's field equations published in 1916. Regardless of how the theories have progressed since that time, the second singularity occurring within the Schwarzchild solution was interpreted as a minimum radius for a body of mass *m*, which apparently led to the formation of a black-hole. Yet both Einstein - who derived the entire theory of general relativity - and Karl Schwarzchild - who originally solved the field equations *exactly* - both explicitly denied the existence of singularities in physical reality. In fact, the Schwarzchild solution in its original form *does not hold* unless the 'discontinuity' alluded to in relation to black holes is forced into the origin at $r = 0$. All of these facts, however, seem to be of little importance to those who have made it their life's work to pursue an avenue of research that is based entirely upon a mathematical expression that has no physical basis. They claim that these objects *must* exist because they are given by Einstein's general relativity but, when it is shown that the man who derived general relativity itself says they are not real, he must be ignored! How can an object be said to exist on the basis of a scientific theory, when the creator of that theory states adamantly that such objects are simply a product of the mathematical model employed.

An adequate explanation of what a 'singularity' in physical space, or space-time, is does not seem to exist. It is described as the point at which the curvature of space-time becomes infinite and the known laws of physics break down but, in reality, it seems that a singularity within general relativity is exactly the same as a singularity within any other theory - meaningless as a physical entity. This is not to say that general relativity is incorrect due to the existence of this singularity within the mathematics of the theory but, rather, that the existence of these singularities demonstrates possible limits to Einstein's theory, and that some deductions from that theory, following work carried out in the 1950s and 60s, may be in error. Hence, as was suggested earlier, it seems sensible for this entire area to be viewed again without any preconceived ideas as to the final answer. Only in this way will satisfactory answers emerge and physics be allowed to progress safely in pursuit of the truth.

**References:**

[1] Oxford Dictionary of Physics Fifth Edition 2005, Oxford University Press, Oxford, 2005

[2] J. Michell, Philos. Trans. Royal Soc., **74**, 35, 1784

[3] J-L. Tassoul & M. Tassoul, *'A Concise History of Solar and Stellar Physics'*, Princeton University Press, Princeton NJ, 2004

4] K. Schwarzchild, *'On the gravitational field of a mass point according to Einstein's theory'*, Sitzungsberichte der Koniglich Preussischen Akademie der Wissenschaften zu Berlin, Phys-Math. Klasse, **189**, 1916,
(Translated by S. Antoci & A. Loinger, arXiv:physics/9905030)

[5] S.E. Bloomer & J. Dunning-Davies, *'Blackholes, other exotic stars and conventional wisdom'* , Apeiron Vol 12, 3, 2005 (http://redshift.vif.com/Apeiron_Home.htm)

[6] R. Oppenheimer & H. Snyder, Phys. Rev. **56**, 455, 1939

[7] R. Adler, M. Bazin, M. Schiffer, *'Introduction to General relativity,'* McGraw-Hill, New York, 1965

[8] J. Dunning-Davies, *'Exploding a myth, conventional wisdom or scientific truth?'*, Horwood Publishing Ltd., Chichester, 2007

[9] P.A.M Dirac, *'General Theory of Relativity',* Princeton University Press, Princeton NJ, 1996

[10] A. Einstein, *Annals of Mathematics*, **40**, 922, (1939)

[11] R.M. Wald, *'General Relativity',* University of Chicago Press, Chicago, 1984